\documentstyle[12pt]{article}
\def\bar{\overline}

\def\1{{\chi}}

\begin{document}
\title {{\bf Remarks on the sequential effect algebras}\thanks{This project is supported by Natural Science
Found of China (10771191 and 10471124).}}
\author {Shen Jun$^{1,2}$, Wu Junde$^{1}$\date{}\thanks{Corresponding author: wjd@zju.edu.cn}}
\maketitle $^1${\small\it Department of Mathematics, Zhejiang
University, Hangzhou 310027, P. R. China}

$^2${\small\it Department of Mathematics, Anhui Normal University,
Wuhu 241003, P. R. China}

\begin{abstract} {In this paper, first, we answer affirmatively an open problem which was presented in 2005 by professor
Gudder on the sub-sequential effect algebras. That is, we prove that
if $(E,0,1, \oplus , \circ)$ is a sequential effect algebra and $A$
is a commutative subset of $E$, then the sub-sequential effect
algebra $\bar{A}$ generated by $A$ is also commutative. Next, we
also study the following uniqueness problem: If $na=nb=c$ for some
positive integer $n\geq 2$, then under what conditions $a=b$ hold?
We prove that if $c$ is a sharp element of $E$ and $a|b$, then
$a=b$. We give also two examples to show that neither of the above
two conditions can be discarded.}

\end{abstract}


\vskip0.2in

{\bf 1. Introduction}

\vskip 0.2 in

Effect algebra is an important logic model for studying quantum
effects or observations which may be {\it fuzzy} or {\it unsharp}
(see [1]), to be precise, an effect algebra is a system $(E,0,1,
\oplus)$, where 0 and 1 are distinct elements of $E$ and $\oplus$ is
a partial binary operation on $E$ satisfying:

(EA1) If $a\oplus b$ is defined, then $b\oplus a$ is defined and
$b\oplus a=a\oplus b$.

(EA2) If $a\oplus (b\oplus c)$ is defined, then $(a\oplus b)\oplus
c$ is defined and $$(a\oplus b)\oplus c=a\oplus (b\oplus c).$$

(EA3) For each $a\in E$, there exists a unique element $b\in E$ such
that $a\oplus b=1$.

(EA4) If $a\oplus 1$ is defined, then $a=0$.

\vskip 0.1 in

In an effect algebra $(E,0,1, \oplus)$, if $a\oplus b$ is defined,
we write $a\bot b$. For each $a\in E$, it follows from (EA3) that
there exists a unique element $b\in E$ such that $a\oplus b=1$, we
denote $b$ by $a'$. Let $a, b\in E$, if there exists an element
$c\in E$ such that $a\bot c$ and $a\oplus c=b$, then we say that
$a\leq b$ and write $c=b\ominus a$. It follows from [1] that $\leq $
is a partial order of $(E,0,1, \oplus)$ and satisfies that for each
$a\in E$, $0\leq a\leq 1$, $a\bot b$ if and only if $a\leq b'$.

\vskip 0.1 in

Let $(E,0,1, \oplus)$ be an effect algebra and $a\in E$. If $a\wedge
a'=0$, then $a$ is said to be a {\it sharp element} of $E$. The set
$E_s=\{x\in E|\ x\wedge x'=0\}$ is called the set of all sharp
elements of $E$ (see [2-3]).

\vskip 0.1 in

As we knew, two measurements $a$ and $b$ cannot be performed
simultaneously in general, so they are frequently executed
sequentially ([4]). We denote by $a\circ b$ a sequential measurement
in which $a$ is performed first and $b$ second and call $a\circ b$ a
{\it sequential product} of $a$ and $b$.  Thus, it is an important
and interesting project to study effect algebras which have a
sequential product $\circ$ with some nature properties. To be
precise:

\vskip 0.1 in

A {\it sequential effect algebra} (SEA) is an effect algebra
$(E,0,1, \oplus)$ and another binary operation $\circ $ defined on
$(E,0,1, \oplus)$ satisfying [5]:

(SEA1) The map $b\mapsto a\circ b$ is additive for each $a\in E$,
that is, if $b\bot c$, then $a\circ b\bot a\circ c$ and $a\circ
(b\oplus c)=a\circ b\oplus a\circ c$.

(SEA2) $1\circ a=a$ for each $a\in E$.

(SEA3) If $a\circ b=0$, then $a\circ b=b\circ a$.

(SEA4) If $a\circ b=b\circ a$, then $a\circ b'=b'\circ a$ and for
each $c\in E$, $a\circ (b\circ c)=(a\circ b)\circ c$.

(SEA5) If $c\circ a=a\circ c$ and $c\circ b=b\circ c$, then
$c\circ(a\circ b)=(a\circ b)\circ c$ and $c\circ(a\oplus b)=(a\oplus
b)\circ c$ whenever $a\bot b$.

\vskip 0.1 in

Let $(E,0,1, \oplus, \circ)$ be a sequential effect algebra. If $a,
b\in E$ and $a\circ b=b\circ a$, then we say $a$ and $b$ is {\it
sequentially independent} and denoted by $a|b$.

\vskip 0.1 in

{\bf Lemma 1 ([1, 5]).} If $(E,0,1, \oplus, \circ)$ is a sequential
effect algebra and $a,b,c\in E$, then

(1) $a\perp b$, $a\perp c$ and $a\oplus b=a\oplus c$ implies that
$b=c$.

(2) $a\in E_s$ if and only if $a\circ a=a$.

(3) If $c\in E_s$, then $a\leq c$ if and only if $a=a\circ c=c\circ
a$.

\vskip 0.2 in

{\bf 2. Sub-sequential effect algebra generated by a subset}

\vskip 0.2 in

Let $(E,0,1, \oplus, \circ)$ be a sequential effect algebra and $F$
a nonempty subset of $E$. We call $F$ a {\it{sub-sequential effect
algebra}} of $(E,0,1, \oplus, \circ)$ if $0,1\in F$ and $(F,0,1,
\oplus, \circ)$ itself is a sequential effect algebra. From the
definition of sub-sequential effect algebra, it is easy to see that
a nonempty subset $F$ of $(E,0,1, \oplus, \circ)$ is a
sub-sequential effect algebra if and only if $F$ is closed under all
the three operations $\oplus$, $\circ$ and $'$. Moreover, if $A$ is
a nonempty subset of $E$, it is easy to see that there exists a
smallest sub-sequential effect algebra $\bar{A}$ of $E$ which
contains $A$ (That is, the intersection of all sub-sequential effect
algebras containing $A$). We call $\bar{A}$ {\it{the sub-sequential
effect algebra generated by $A$}}. In 2005, Professor Gudder
presented the following open problem (see [6, Problem 17]):

\vskip 0.1 in

{\bf Problem 1.} If $(E,0,1, \oplus , \circ)$ is a sequential effect
algebra and $A$ a commutative subset of $E$ (That is, $a|b$ for all
$a, b\in A$), is $\bar{A}$ commutative ?

\vskip 0.1 in

In this paper, we answer the problem affirmatively. That is:

\vskip 0.1 in

{\bf Theorem 1.} Let $(E,0,1, \oplus , \circ)$ be a sequential
effect algebra and $A$ a commutative subset of $(E,0,1, \oplus ,
\circ)$. Then  $\bar{A}$ is also commutative.

{\bf Proof.} Let $\bigwedge=\{F|\ F\ be\ a$ commutative subset of
$E$ containing $A$\}. We order $\bigwedge$ by including. Using
Zorn's Lemma, it is easy to see that there exists a maximal element
$F_0$ in $\bigwedge$. That is, $F_0$ is a maximal commutative subset
of $E$ containing $A$.

We now prove that $F_0$ is a sub-sequential effect algebra of $E$:

If $a\in F_0$, then for each $c\in F_0$, $c|a$, so $c|a'$ by (SEA4).
By maximality, we have $a'\in F_0$.

If $a,b\in F_0$, then for each $c\in F_0$, $c|a$, $c|b$, so
$c|(a\circ b)$ by (SEA5). By maximality, we have $(a\circ b)\in
F_0$.

If $a,b\in F_0$ and $a\perp b$, then for each $c\in F_0$, $c|a$,
$c|b$, so $c|(a\oplus b)$ by (SEA5). By maximality, we have
$(a\oplus b)\in F_0$.

So $F_0$ is closed under all the three operations $\oplus$, $\circ$
and $'$.

Thus, $F_0$ is a sub-sequential effect algebra of $(E,0,1, \oplus ,
\circ)$ containing $A$. Since $\bar{A}$ is the smallest
sub-sequential effect algebra of $(E, 0, 1, \oplus, \circ)$
containing $A$, we have $\bar{A}\subseteq F_0$ and $\bar{A}$ is also
commutative.

\vskip 0.1 in

Moreover, for general subset $A$ of $E$, we can describe the
structure of $\bar{A}$, that is

\vskip 0.1 in

{\bf Theorem 2.} Let $(E,0,1, \oplus , \circ)$ be a sequential
effect algebra and $A$ a subset of $E$. If we denote

$A_1=A\bigcup (\bigcup\limits_{a\in A}a')\bigcup
(\bigcup\limits_{a,b\in A}a\circ b)\bigcup (\bigcup\limits_{a,b\in
A\ and\ a\perp b}a\oplus b)$,

$A_2=A_1\bigcup (\bigcup\limits_{a\in A_1}a')\bigcup
(\bigcup\limits_{a,b\in A_1}a\circ b)\bigcup (\bigcup\limits_{a,b\in
A_1\ and\ a\perp b}a\oplus b)$,

$\cdots$

$A_n=A_{n-1}\bigcup (\bigcup\limits_{a\in A_{n-1}}a')\bigcup
(\bigcup\limits_{a,b\in A_{n-1}}a\circ b)\bigcup
(\bigcup\limits_{a,b\in A_{n-1}\ and\ a\perp b}a\oplus b)$,

$\cdots$

$\Gamma=\bigcup\limits_{n=1}\limits^{\infty}A_n$.

Then $\bar{A}=\Gamma$.

{\bf Proof.} First we prove that $\Gamma$ is a sub-sequential effect
algebra of $(E,0,1, \oplus , \circ)$.

If $a\in \Gamma$, then $a\in A_n$ for some $n$, so $a'\in
A_{n+1}\subseteq \Gamma$.

If $a,b\in \Gamma$, then $a,b\in A_n$ for some $n$, so $(a\circ
b)\in A_{n+1}\subseteq \Gamma$.

If $a,b\in \Gamma$ and $a\perp b$, then $a,b\in A_n$ for some $n$,
so $(a\oplus b)\in A_{n+1}\subseteq \Gamma$.

Thus, $\Gamma$ is closed under all the three operations $\oplus$,
$\circ$ and $'$. So $\Gamma$ is a sub-sequential effect algebra of
$(E, 0, 1, \oplus, \circ)$.

Of course $A\subseteq\Gamma$. Since $\bar{A}$ is the smallest
sub-sequential effect algebra of $(E, 0, 1, \oplus, \circ)$
containing $A$, we have $\bar{A}\subseteq \Gamma$. On the other
hand, by induction, it is easy to see that $A_{n}\subseteq \bar{A}$
for all $n$. Thus $\Gamma\subseteq\bar{A}$. So $\Gamma=\bar{A}$.

\vskip 0.1 in

Note that by using Theorem 2 we can also answer professor Gudder's
problem by a constructive way, we omit the process.

\vskip0.2in

{\bf 3. An addition property of sequential effect algebras}

\vskip 0.2 in

Let $(E,0,1, \oplus , \circ)$ be a sequential effect algebra, $a,
b\in E$. If $\underbrace{a\oplus a\cdots\oplus a}\limits_{the\
number\ is\ n}$ is defined, we denote it by $na$. Now, we are
interested in the following uniqueness problem: If for some positive
integer $n_0\geq 2$, $n_0a=n_0b$, then under what conditions $a=b$
hold? We have

\vskip 0.1 in

{\bf Theorem 3.} Let $(E,0,1, \oplus , \circ)$ be a sequential
effect algebra, $a, b\in E$ and for some positive integer $n_0\geq
2$, $n_0a=n_0b=c$. If $c\in E_s$ and $a|b$, then $a=b$.

\vskip 0.1 in

{\bf Proof.} Since $a\leq c$, by Lemma 1, $a=a\circ c$, similarly
$b=b\circ c$.

By (SEA1), we have $a\circ c=a\circ (n_0b)=n_0(a\circ b)$, $b\circ
c=b\circ (n_0a)=n_0(b\circ a)$.

Note that $a|b$, so $a\circ b=b\circ a$ and $a\circ c=b\circ c$.
Thus $a=b$.

\vskip 0.1 in

Now, we show that neither of the two conditions in Theorem 3 can be
discarded.

\vskip 0.1 in

{\bf Example 1.} Let $I_1=[0,1]$, $I_2=[0,1]$, $E=HS(I_1,I_2)$ be
the horizontal sum of $I_1,I_2$ (see [5, Section 8, the Example in
$P_{109}$]). For each $t\in[0,1]$, if it is in $I_1$, we denote it
by $\hat{t}$; if it is in $I_2$, we denote it by $\check{t}$. Let
$a=\hat{\frac{1}{n_0}}$, $b=\check{\frac{1}{n_0}}$. Then
$n_0a=1=n_0b$, $1\in E_s$, $a\neq b$, $a\circ b\neq b\circ a$. So
the condition $a|b$ in Theorem 3 can not be discarded.

\vskip 0.1 in

{\bf Example 2.} Let ${\mathbf{N}}$ be the nonnegative integer set,
$n_0$ be a positive integer and $n_0\geq 2$,
$E_0=\{0,1,a_{n,m},b_{n,m}|\ n,m\in {\mathbf{N}},\ n_0-1\geq m,\
n^2+m^2\neq 0\}$.

First, we define a partial binary operation $\oplus$ on $E_0$ as
follows (when we write $x\oplus y=z$, we always mean $x\oplus
y=z=y\oplus x$):

For each $x\in E_0$, $0\oplus x=x$,

$$
a_{n,m}\oplus a_{r,s}= \left\{
  \begin{array}{ll}
    a_{n+r,m+s}\ , & \hbox{$if\ m+s<n_0$;} \\
    a_{n+r+n_0,m+s-n_0}\ , & \hbox{$if\ m+s\geq n_0$.}
  \end{array}
\right.
$$

$$
a_{n,m}\oplus b_{r,s}= \left\{
  \begin{array}{ll}
    b_{r-n,s-m}\ , & \hbox{$if\ n\leq r,\ m\leq s,\ (r-n)^2+(s-m)^2\neq 0$;} \\
    1\ , & \hbox{$if\ n=r,\ m=s$;} \\
    b_{r-n-n_0,s-m+n_0}\ , & \hbox{$if\ n+n_0\leq r,\ m>s$.}
  \end{array}
\right.
$$

No other $\oplus$ operation is defined.

Next, we define a binary operation  $\circ$  on $E_0$ as follows
(when we write $x\circ y=z$, we always mean $x\circ y=z=y\circ x$):

For each $x\in E_0$, $0\circ x=0$, $1\circ x=x$,

$a_{n,m}\circ a_{r,s}=0$, $a_{n,m}\circ b_{r,s}=a_{n,m}$,

$$
b_{n,m}\circ b_{r,s}= \left\{
  \begin{array}{ll}
    b_{n+r,m+s}\ , & \hbox{$if\ m+s<n_0$;} \\
    b_{n+r+n_0,m+s-n_0}\ , & \hbox{$if\ m+s\geq n_0$.}
  \end{array}
\right.
$$

Now, we prove that $E_0$ is a sequential effect algebra.

In fact, (EA1) and (EA4) are trivial.

We verify (EA2), for simplicity, we omit the trivial cases about
0,1:

$ a_{k,j}\oplus(a_{n,m}\oplus a_{r,s})=(a_{k,j}\oplus a_{n,m})\oplus
a_{r,s}$
$$=\left\{
  \begin{array}{ll}
    a_{k+r+n,s+j+m}\ , & \hbox{$if\ s+j+m<n_0$;} \\
    a_{k+r+n+n_0,s+j+m-n_0}\ , & \hbox{$if\ n_0\leq s+j+m<2n_0$;} \\
    a_{k+r+n+2n_0,s+j+m-2n_0}\ , & \hbox{$if\ s+j+m\geq 2n_0$.}
  \end{array}
\right.
$$

Each $ a_{k,j}\oplus(a_{n,m}\oplus b_{r,s})$ or $(a_{k,j}\oplus
a_{n,m})\oplus b_{r,s}$ is defined if and only if one of the
following four conditions is satisfied, at this case,

$ a_{k,j}\oplus(a_{n,m}\oplus b_{r,s})=(a_{k,j}\oplus a_{n,m})\oplus
b_{r,s}$

$$=\left\{
  \begin{array}{ll}
    b_{r-k-n,s-j-m}\ , & \hbox{$if\ k+n\leq r,\ j+m\leq s,\ (r-k-n)^2+(s-j-m)^2\neq 0$;} \\
    b_{r-k-n-n_0,s-j-m+n_0}\ , & \hbox{$if\ k+n+n_0\leq r,\ s<j+m\leq n_0+s,$} \\
      & \hbox{~~~~~~~~$(r-k-n-n_0)^2+(s-j-m+n_0)^2\neq 0$;} \\
    b_{r-k-n-2n_0,s-j-m+2n_0}\ , & \hbox{$if\ k+n+2n_0\leq r,\ n_0+s<j+m$;} \\
    1\ , & \hbox{$if\ (r-k-n)^2+(s-j-m)^2=0\ or$} \\
      & \hbox{~~~~~~~~$(r-k-n-n_0)^2+(s-j-m+n_0)^2=0$.}
  \end{array}
\right.
$$ Thus, (EA2) is hold.

(EA3) is clear since $a_{n,m}\oplus b_{n,m}=1$. Thus, $(E_0,0,1,
\oplus)$ is an effect algebra.

Moreover, we verify that $(E_0,0,1, \oplus , \circ)$ is a sequential
effect algebra.

(SEA2) and (SEA3) and (SEA5) are trivial.

We verify (SEA1), for simplicity, we omit the trivial cases about
0,1:

$ a_{k,j}\circ(a_{n,m}\oplus a_{r,s})=a_{k,j}\circ a_{n,m}\oplus
a_{k,j}\circ a_{r,s}=0$.

$$
b_{k,j}\circ(a_{n,m}\oplus a_{r,s})=b_{k,j}\circ a_{n,m}\oplus
b_{k,j}\circ a_{r,s}= \left\{
  \begin{array}{ll}
    a_{n+r,m+s}\ , & \hbox{$if\ m+s<n_0$;} \\
    a_{n+r+n_0,m+s-n_0}\ , & \hbox{$if\ m+s\geq n_0$.}
  \end{array}
\right.
$$

When $a_{n,m}\oplus b_{r,s}$ is defined,

$ a_{k,j}\circ(a_{n,m}\oplus b_{r,s})=a_{k,j}\circ a_{n,m}\oplus
a_{k,j}\circ b_{r,s}=a_{k,j}$,

$ b_{k,j}\circ(a_{n,m}\oplus b_{r,s})=b_{k,j}\circ a_{n,m}\oplus
b_{k,j}\circ b_{r,s}$

$$=\left\{
  \begin{array}{ll}
    b_{r+k-n,s+j-m}\ , & \hbox{$if\ n\leq r,\ m\leq s,\ j+s<n_0+m$;} \\
    b_{r+k-n,s+j-m}\ , & \hbox{$if\ n+n_0\leq r,\ s<m\leq j+s$;} \\
    b_{r+k-n+n_0,s+j-m-n_0}\ , & \hbox{$if\ n\leq r,\ n_0+m\leq j+s$;} \\
    b_{r+k-n-n_0,s+j-m+n_0}\ , & \hbox{$if\ n+n_0\leq r,\ j+s<m$.}
  \end{array}
\right.
$$ Thus, (SEA1) is true.

We verify (SEA4), for simplicity, we omit also the trivial cases
about 0,1:

$ a_{k,j}\circ(a_{n,m}\circ a_{r,s})=(a_{k,j}\circ a_{n,m})\circ
a_{r,s}=0$.

$ a_{k,j}\circ(a_{n,m}\circ b_{r,s})=(a_{k,j}\circ a_{n,m})\circ
b_{r,s}=0$.

$ a_{k,j}\circ(b_{n,m}\circ b_{r,s})=(a_{k,j}\circ b_{n,m})\circ
b_{r,s}=a_{k,j}$.

$ b_{k,j}\circ(b_{n,m}\circ b_{r,s})=(b_{k,j}\circ b_{n,m})\circ
b_{r,s}$
$$=\left\{
  \begin{array}{ll}
    b_{k+r+n,s+j+m}\ , & \hbox{$if\ s+j+m<n_0$;} \\
    b_{k+r+n+n_0,s+j+m-n_0}\ , & \hbox{$if\ n_0\leq s+j+m<2n_0$;} \\
    b_{k+r+n+2n_0,s+j+m-2n_0}\ , & \hbox{$if\ s+j+m\geq 2n_0$.}
  \end{array}
\right.
$$ Thus (SEA4) is hold and $(E_0,0,1, \oplus , \circ)$ is a sequential effect algebra.

Finally, we show that the condition $c\in E_s$ in Theorem 3 can not
be discarded.

Indeed, since $a_{n,0}\oplus a_{r,0}=a_{n+r,0}$, so
$n_0a_{1,0}=a_{n_0,0}$. Note that
$$
a_{0,m}\oplus a_{0,s}= \left\{
  \begin{array}{ll}
    a_{0,m+s}\ , & \hbox{$if\ m+s<n_0$;} \\
    a_{n_0,m+s-n_0}\ , & \hbox{$if\ m+s\geq n_0$.}
  \end{array}
\right.
$$
Thus, $(n_0-1)a_{0,1}=a_{0,n_0-1}$, $n_0a_{0,1}=(n_0-1)a_{0,1}\oplus
a_{0,1}=a_{0,n_0-1}\oplus a_{0,1}=a_{n_0,0}$, that is,
$n_0a_{1,0}=a_{n_0,0}=n_0a_{0,1}$. Note that $a_{n_0,0}\circ
a_{n_0,0}=0$, so $a_{n_0,0}\not\in (E_0)_s$.

\vskip0.3in

\centerline{\bf References}

\vskip0.2in

\noindent [1]. D. J. Foulis and M. K. Bennett: Effect algebras and
unsharp quantum logics. {\em Found. Phys. \/} {\bf 24}, 1331(1994).

\noindent [2]. S. Gudder: Sharply dominating effect algebras. {\em
Tatra Mt. Math. Publ.\/} {\bf 15}, 23(1998).

\noindent [3]. Z. Riecanova and J. D. Wu: States on sharply
dominating effect algebras. {\em Sci. in China A: Math.\/} {\bf 51},
907(2008).

\noindent [4]. S. Gudder and G. Nagy: Sequential quantum
measurements. {\em J. Math. Phys.\/} {\bf 42}, 5212(2001).

\noindent [5]. S. Gudder and R. Greechie: Sequential products on
effect algebras. {\em Rep. Math. Phys.\/} {\bf 49}, 87(2002).

\noindent [6]. S. Gudder: Open problems for sequential effect
algebras. {\em Inter. J. Theory. Phys.\/}  {\bf 44}, 2219(2005).

\end{document}